\documentstyle[12pt]{article}
\def\baselinestretch{1.2}
\newcommand \ca {{\cal A}}
\newcommand \cb {{\cal B}}
\newcommand \cf {{\cal F}}
\newcommand \ch {{\cal H}}
\newcommand \cg {{\cal G}}
\newcommand \chh {\hat {\cal H}}
\newcommand \cm {{\cal M}}
\newcommand \ct {{\cal T}}
\newcommand \cu {{\cal U}}
\newcommand \hphi {{\hat \phi}}
\newcommand \hp {{\hat p}}

\begin{document}

\title{
Generalized Hybrid Monte-Carlo
}

\author{
{\bf \underline {Ra\'ul Toral}} \\
Departament de F\'\i sica,\\
Universitat de les Illes Balears\\
07071-Palma de Mallorca, Spain.\\
{\bf A.L. Ferreira}\\
Departamento de F\'\i sica\\
Universidade de Aveiro \\
3800 Aveiro, Portugal
}
\maketitle

\begin{abstract}
We propose a modification of the Hybrid Monte-Carlo method
to sample equilibrium distributions of continuous field models. The
method allows an efficient implementation of Fourier acceleration and
is shown to reduce completely critical slowing down for the
Gaussian model, i. e., $z=0$.
\end{abstract}

\newpage

The development of efficient numerical algorithms to study equilibrium
properties of field-theoretical models near second order phase
transitions is a very important subject in particle and statistical
physics\cite{Metropolis,sokal1,sokal2,kennedy,wolff}.
For that purpose, several methods such as Molecular Dynamics, Langevin
and Monte--Carlo (MC) have been used. While the first two methods suffer
from systematic step--size time discretization errors which may affect
the computed mean values of observables, the only errors present in MC
methods are of statistical origin and can be easily controlled, in
principle, by varying the number of samplings. In practice, however, it
turns out that for many problems of interest the number of
configurations necessary to achieve a given small error is very large
and grows as some power of the system size, thus requiring too much
computer time.

In its simplest form, MC introduces a stochastic dynamics which involves
the proposal of a random new local field configuration plus an
acceptance/rejection step. This method can be inefficient to implement
in some systems. The Hybrid Monte Carlo (HMC) algorithm first proposed
by Duane {\it et al.} \cite{hybrid}, uses Molecular Dynamics to propose
new configurations then requiring a single acceptance/rejection step
every time the whole system is updated, which is a substantial
improvement over other MC methods. The usual implementation of HMC
relies on an appropriate numerical integration of the corresponding
Hamiltonian dynamics of the system.

In this paper, we present a generalization of HMC which is based upon
the numerical integration of a non-hamiltonian dynamics which, however,
conserves the energy of the system\cite{toral}. Our method turns out to
be closely related to the use of a time-step matrix in a Langevin
integration scheme \cite{batrouni} and thus can be regarded as an exact
version of this method in the sense of not being affected by step-size
errors. We first define our generalized algorithm showing that it obeys
correctly the detailed balance condition. As an application, we study
the Gaussian model. The simplicity of the model enables us to carry a
systematic analytic study of correlation times. We also show that the
optimal implementation of our method for the Gaussian model is actually
a Fourier acceleration scheme which completely reduces critical slowing
down (CSD).

CSD theory\cite{hh} tells us that near second--order
phase transitions the correlation time, $\tau_{\hat O}$, of a measured
observable $\hat O$, increases as $\tau_{\hat O} \sim \xi^z$, being
$\xi$ the correlation length and $z$ the dynamical critical exponent.
$\tau_{\hat O}$ can be defined as some measure of the relaxation time of
the correlation function  of the observable $\hat O$:
\begin{equation}
C_{\hat O}(t) = \frac{<\hat O(t) \hat O(0)>-<\hat O(t)>< \hat O(0)>}
          {<\hat O^2(0)> - <\hat O(0)>^2}
\end{equation}
In a numerical simulation of a d--dimensional system the computer time,
$T_{\hat O}$, required to measure the observable $\hat O$ at a given
error behaves as $T_{\hat O} \equiv t_{\hat O} L^d \sim L^{d+c} (2
\tau_{\hat O} + 1) $. The factor $L^d$ is always present as we simulate
a system of size $N=L^d$ as a consequence of the increase of the number
of degrees of freedom. The factor $L^c$, present in HMC, is a
consequence of the fact that simulations at a constant correlation
length at bigger sizes may require an additional computer effort in
order to keep the acceptance probability within reasonable limits.
The last factor $(2 \tau_{\hat
O} + 1)$ takes into account the number of effectively independent
configurations produced by the algorithm. For finite systems close
enough to the critical point, the correlation time $\tau_{\hat O}$
increases with system side as $\tau_{\hat O} \sim L^z$. For the local
updating schemes such as heat-bath or Metropolis the exponent $z$ being
near 2 strongly demands on  computer time (although the above defined
exponent $c$ is actually $0$ for these algorithms). For spin models the
collective updating scheme of Swendsen and Wang \cite{sw} has proven
quite successful in reducing the dynamical critical exponent and
overcoming CSD. For continuous models a Multigrid Monte-Carlo
Method\cite{mgrid} was proposed which can also reduce CSD in certain
cases. Also the time-step matrix Langevin method\cite{batrouni} can be
helpful for some models if an appropriate matrix is chosen. The last two
algorithms were shown to reduce the exponent $z$ for the Gaussian model
but it is not clear if some reducing can be achieved in interacting
models such as the $\phi^4$ model. The standard implementation of HMC
was not observed to ameliorate CSD for the $\phi^4$ model\cite{mehlig}
and for the Gaussian model it was shown \cite{pend} that the algorithm
reduces $z=1$ if an appropriate tuning of the trajectory length is
taken. For an application of the standard HMC to study the two
dimensional XY model see \cite{gupta2}. Yet another algorithm also used
for simulation of these models is Overrelaxation (see \cite{kennedy} and
references therein) which was also shown to produce $z=1$ for the
Gaussian model\cite{wolff2}. The modified HMC we propose allows to
reduce $z=0$ for the Gaussian model.

We now describe the generalized HMC algorithm. Let us consider a system
\mbox{$(\phi_1, \phi_2, \dots, \phi_N) \equiv [\phi]$}, of $N=L^d$
scalar variables, whose statistical equilibrium properties are defined
through the Gibbs factor $\exp(-\ch)$, by its Hamiltonian $\ch[{\phi}]$.
The variables $[\phi]$ are considered to be generalized coordinates and
a set of conjugate  momenta $(p_1, p_2, \dots, p_N) \equiv [p] $
associated with a kinetic energy $\ch_K = \sum_{i=1}^N p_i^2/2$ is
introduced. The variable $p_i$ can be in general a vector variable with
$D$ components, $p_i=(p^1_i,p^2_i,\dots, p^D_i)$. The total Hamiltonian
is $\chh= \ch+\ch_K$. We propose the following dynamics:
\begin{eqnarray}
\frac{d\phi_i}{dt} &=& \sum_{s=1}^D \sum_{j=1}^N (\ca^s)_{ij}
p^s_j \\
\nonumber
\frac{d p^s_i}{dt} &=&  \sum_{j=1}^N(\ca^s)_{ji} F_j ~~~~~~~,s=1,\dots, D
\end{eqnarray}
or, written in more compact vector notation:
\begin{eqnarray}
\label{baseq}
\frac{d\phi}{dt} &=& \sum_{s=1}^D \ca^s p^s \\
\nonumber
\frac{d p^s}{dt} &=&  (\ca^s)^T F ~~~~~~~,s=1,\dots, D
\end{eqnarray}
where the $\ca^s$ are some linear operators which can be
represented as a matrix,  and $F_j$ represents the force as computed
from the Hamiltonian $-\frac{\partial}{\partial \phi_j} \ch$.
The standard HMC substitutes the above dynamics by the Hamiltonian
dynamics which can be obtained from the above set of equations by
considering $D=1$ and $\ca$ equal to the identity operator.

It is an essential property that can be easily verified that the proposed
dynamics in equations (\ref{baseq}) exactly conserves energy, i.e.,
$d\chh / dt=0$. For the approximate integration of the previous
equations of motion the ``leap--frog'' scheme can be used, introducing a
discrete mapping $ [\phi(t),p(t)] \rightarrow  [\phi(t+\delta
t),p(t+\delta t)] = G^{\delta t}([ \phi(t),p(t)])$ of phase space,
dependent on the time step $\delta t$ chosen. The total energy, as a
result of the time discretization used in the leap--frog scheme, is no
longer conserved and its variation can be controlled by varying $\delta
t$. The leap-frog approximation reads:
\begin{eqnarray}
\label{bceq}
\phi' &=& \phi + \delta t \sum_{s=1}^D \ca^s p^s + \frac{(\delta t)^2}{2}
\sum_{s=1}^D \ca^s (\ca^s)^T F([\phi])  \\
\nonumber
p'^s &=& p^s + \frac{\delta t}{2} (\ca^s)^T \left( F([\phi]) + F([\phi'])
\right)
\end{eqnarray}
We define yet another mapping obtained from n iterations of the previous
mapping, i.e. $\cg = (G^{\delta t})^n$. The configuration obtained when
one applies $\cg$ is then accepted or rejected in such a way that detailed
balance is verified in order to sample the canonical distribution for
the Hamiltonian $\ch$. As in the standard HMC the momenta variables are
refreshed after every acceptance/rejection step according to the
Gaussian distribution of independent variables $\exp(-\ch_K)$. The
evolution given by $\cg$ ($n$ leap--frog steps) and the acceptance/rejection
step constitute what is called 1 MC trial. Detailed balance is obeyed
if one requires $G^{\delta t}$ to be time reversible and
area preserving and if one
accepts the new configuration with probability
\mbox{$\min[1,\exp(-\Delta \chh )]$}, where \mbox {$\Delta \chh =
\chh(\cg([\phi,p])) - \chh([\phi,p])$}. The time reversibility $G^{\delta
t}([\phi',-p'])= [\phi,-p]$ can be easily verified to be obeyed by
equations (\ref{bceq}). One can also proof that the area preserving
property is verified by these equations. The above properties are
satisfied for arbritrary matrices $\ca^s$ provided the associated mapping
$G^{\delta t}$ remains a one to one mapping in phase-space. We have
defined a variety of HMC--type methods characterized by a particular choice of
matrices $\ca^s$. One can then choose the matrices $\ca^s$ that better
suit a particular problem. We have shown before how a particular version
of the above generalized HMC can be used to simulate conserved order
parameter systems\cite{pwork}.

We compare now our method with the one introduced in reference
\cite{batrouni} based upon the numerical integration of a Langevin
equation using a matrix time--step. This method is based upon the
observation that the stationary probability distribution of the Langevin
equation:
\begin{equation}
\frac{\partial \phi_i(\tau) }{\partial \tau} = - \frac{\delta \ch}{\delta
\phi_i} + \sqrt{2} \xi_i(\tau)
\end{equation}
is precisely $\exp(-\ch)$. Here $\xi_i(\tau)$ are stochastic Gaussian
random variables of mean zero and correlations
$<\xi_i(\tau)\xi_j(\tau')> = \delta_{ij}\delta(\tau-\tau')$. The
solution of the equation is approximated by:
\begin{equation}
\phi_i(\tau+\delta \tau)=\phi_i(\tau)+
\sum_j\left[-\delta \tau \epsilon_{ij}\frac{\delta \ch}{\delta
\phi_j}+\sqrt{2\delta \tau} \sqrt{\epsilon}_{ij} \eta_j \right]
\end{equation}
Where $\epsilon_{ij}$ is an arbitrary matrix and  $\eta_j$ is a
Gaussian variable of mean zero and correlations $<\eta_i\eta_j>=
\delta_{ij}$. This corresponds exactly to the one step leap--frog
approximation of the generalized HMC introduced above (equation
(\ref{bceq}) with $D=1$) if we identify: $(\delta t)^2/2=\delta \tau$
and $\ca \ca^T=\epsilon$. The main difference between the two methods
is the presence of an acceptance/rejection step in the generalized HMC
absent in the numerical integration of the Langevin equation. In that
sense, we can say that the generalized HMC method introduced in this
paper makes exact (in the sense that averages are not biased by the
choice of the time step) the numerical integration of the Langevin
equation using a matrix time step introduced in reference
\cite{batrouni}.

As an application we have considered the Gaussian model defined by the
following Hamiltonian:
\begin{equation}
\ch = \sum_{i=1}^N \left[ \frac{\mu}{2}\phi_i^2 +
\frac{1}{2}\mid\vec \nabla_L\phi_i\mid^2 \right]
\end{equation}
index $i$ runs over the $N=L^2$ sites of a 2-dimensional square lattice,
with periodic boundary conditions (a similar analysis can be carried out
in any spatial dimension but we refer to the case d=2 for simplicity).
$\vec \nabla_L$ is the usual lattice discretized version of the gradient
operator. This problem can be better analyzed in Fourier space. The
total Hamiltonian $\chh$ in terms of the Fourier transform of fields and
momenta space is:
\begin{equation}
\chh = \sum_{k=1}^N \left[ \frac{\omega_k^2}{2} |\hphi_k|^2 +
\frac{1}{2}|\hp_k|^2 \right]
\end{equation}
where $\omega_k$ is given by $\omega_k^2=\mu + 4 \bigl (\sin^2(k_x/2)
+ \sin^2(k_y/2)\bigr )$ and $\hphi_k$ and $\hp_k$ stand for the fields
and momenta variables in Fourier space. We choose the number of
momenta variables associated to a given field equal to 1, $D=1$.
Suppose that we choose for the  matrix $\ca$, generating the dynamics,
a diagonal matrix in Fourier space. Then after $n$ leap-frog steps,
equation (\ref{bceq}) implies that:
\begin{equation}
\left[ \begin{array} {c}
                     \omega_k\hphi_k(n \delta t)\\
                     \hp_k(n \delta t)
        \end{array}\right]                = M_k^n
        \left[ \begin{array}{c}
                     \omega_k\hphi_k(0)\\
                     \hp_k(0)
        \end{array} \right ]
\label{iter}
\end{equation}
for $k=1,\dots,N$. Matrices $M_k^n$ are given by:
\begin{equation}
M_k^n = \left[ \begin{array} {cc}
\cos( n \theta_k) &  \sin( n \theta_k)/\cos(\theta_k/2) \\
-\cos(\theta_k/2)\sin( n \theta_k) & \cos( n \theta_k)
\end{array} \right]
\end{equation}
where we have introduced
$\theta_k = \cos^{-1}(1 - c_k^2/2)$
and $c_k = \hat A_k \omega_k \delta t$ and $\hat A_k$  denoting the
diagonal elements of the matrix $\ca$ in Fourier space. In this model
the different modes evolve independently of each other and the
evolution equations are linear (this is similar to the standard HMC
with $\ca=1$\cite{pend}). For the stability of the leap--frog
integration, the eigenvalues of $M_k$ should lie on the unit circle of
the complex plane which happens to be the case if $c_k$ is between 0
and 2.

By using the evolution equations together with the assumption that the
field variables $\hphi_k(0)$ are in thermal equilibrium and,
therefore, follow the distribution $\exp(-\ch)$, one can compute the
equilibrium average discretization error as:
\begin{equation}
\label{der}
<\chh(n\delta t)-\chh(0)>\equiv<\Delta \chh> = \sum_{k=1}^N
\frac{c_k^4}{32-8c_k^2} \sin^2(n \theta_k)
\label{energy_error}
\end{equation}

In reference \cite{gupta} it was shown that, to a good approximation,
the average acceptance probability $<p_A>$ is related to the average
discretization error $<\Delta \chh>$ by:
\begin{equation}
\label{pa}
<p_A> = {\rm erfc}(\frac{1}{2} \sqrt{<\Delta \chh>})
\label{p_acc}
\end{equation}

We now turn to the question of the optimal choice for the matrix
$\ca$. From equation (\ref{iter}) it is immediately seen that if we
choose the matrix $\ca$ such that $\hat A_k = 1/\omega_k$ the
iteration equations get independent of the mass $\mu$ and all the
modes are equally updated.
This is, in effect, an exact implementation of the method
of Fourier acceleration.
This choice of the matrix clearly reduces
completely CSD ($z=0$) in the sense that correlation times are
independent of the mass even when the mass goes to zero and the model
becomes critical. The standard HMC corresponds here to the choice
$\hat A_k=1$, independent of $k$ \cite{pend}.

Let us compute the computational effort needed to achieve a given
statistical error.  In order to make further analytical calculations, it
is convenient to introduce a set of random variables $\sigma_m$ which
take the value $1$ or $0$ if the configuration proposed after $m$ MC
trials has been accepted or not, respectively. Using this variable we
can write an expression for the field variable after $m$ MC trials,
$\hphi_k(m~ n\delta t)$, as
\begin{eqnarray}
\label{iter2}
\nonumber
\omega_k\hphi_k(m~n\delta t) & = & \sigma_m \left[\cos(n\theta_k)
\omega_k\hphi_k((m-1)~n\delta t) + \frac{\sin(n\theta_k)}{\cos(\theta_k/2)}
\hp_k((m-1)~n\delta t)\right] + \\
{}~  & ~  & (1-\sigma_m) \omega_k\hphi_k((m-1)~n\delta t)
\end{eqnarray}
The momenta $\hp_k(m~n\delta t)$ are the independent random variables
following a Gaussian distribution which are drawn after the m-th
acceptance/rejection step. This equation can be iterated to obtain
$\hphi_k(m~n\delta t)$ in terms of $\hphi_k(0)$ and all the momenta
generated during the evolution.

The variables $\sigma_m$ are Bernoulli variables with probability of
being equal to $1$ equal to the acceptance probability,
$\min[1,\exp(-\Delta \chh(m)]$. This probability depends on the total
change in energy at the $m^{th}$ MC trial, $\Delta \chh(m)$ which is a
function of the initial field configuration and of the momenta generated
at the $j^{th}$ MC trial and variables $\sigma_j$ such that $j<m$. To
proceed we make the approximation that the $\sigma$ variables are all
independently distributed variables with the probability of being equal
to one equal to the average acceptance probability $p=<p_A>$, given by
(\ref{p_acc}). This means that we consider the probability to accept or
reject the whole configuration at a given step to be independent of the
previous ``time-history'' of the system. The approximation is reasonably
good as we will see later. Within this approximation, the correlation
function for the magnetization is:
\begin{equation}
\label{CM}
\bar C_M(m) \equiv C_M(m~ n\delta t)  =
\left[ 1-2p\sin^2(\frac{n\theta_0}{2})\right]^m
\end{equation}

The obtained correlation function is exponential with a correlation time
equal to $\tau_M = -1 / \log(|\bar C_M(1)|)$ in units of MC trials. The
relaxation time of other modes is obtained replacing in the previous
equation $\theta_0$ by $\theta_k$. For the optimal matrix,
$\theta_k\equiv \theta$ is
independent of $k$ and so all the modes relax in the same way.

In figure 1 we compare $\bar C_M(1)$ given by the above correlation function
for the magnetization with simulation results for the case $L=32$ and
$n=4$ as a function of $\delta t$.
The excellent agreement between the analytical
expression and the simulation results shows that the previous
approximation works extremely well for the calculation of the
correlation function for the magnetization. The agreement between
simulation and this approximation actually improves with increasing
system size. We note the following
features (see figure 1): the value of $\tau_M$ as a function of $\delta
t$ for a given $n$ shows a minimum for small values of $\delta t$ and
then $n-1$ zeros for n odd and $n$ zeros for n even. This minimum
disappears for $n$ larger than a given value. We discuss now how
the parameters $n$ and $\delta t$ should be optimized in order to
minimize the computer effort for the magnetization $t_M=(2 \tau_M +1)
n$.

In principle, since there is always a value of $\delta t$ that yields
$\tau_M=0$ for $n=2$,these are obviously the optimal choices. The
corresponding computational effort $t_M=2$ is independent of the system
size corresponding to an exponent $c=0$. However, a closer look shows
that as the system size grows the precise $\delta t$ value at which the
correlation time is zero is very difficult to locate in the sense that a
slight error in the chosen value puts the system in a region of high
(negative) correlations (see figure 1). Furthermore the corresponding
correlation times of the energy would  not be optimized, increasing
enormously as the system size grows.
This becomes apparent when one
discusses the correlation function of the energy. The obtained
approximation for the correlation function of the energy is:
\begin{equation}
\label{CH}
\bar C_{\ch}(m) \equiv C_{\ch}(m~n\delta t) =  [ 1-p\sin^2(n \theta) ]^m
\end{equation}
The correlation time is $\tau_{\ch}=-1/\log(\bar C_{\ch}(1))$. In figure
2 we have compared $\bar C_{\ch}(1)$ given by the above correlation
function to simulation results. The agreement here, although still
reasonable, is not as good as it was in the case of the magnetization.
However, one can still obtain precise quantitative conclusions from this
figure. The value of $\bar C_{\ch}(1)$ has no zeros as a function of
$\delta t$ for a given $n$, in contrast to what happens with $\bar
C_M(1)$. It still shows, however, an absolute minimum at a given $\delta
t$ which approaches zero as $n$ is increased. We have used this minimum
in the correlation function of the energy as a function of $\delta t$ to
find the optimal $n$ for a given system size that minimizes the computer
effort $ (2 \tau_{\ch} + 1) n$. Although we have been unable to obtain
an analytical expression for the optimal values for $n$ and $\delta t$,
a numerical study of equations (\ref{der}), (\ref{pa}) and
(\ref{CH}) allows us to conclude that the optimal value for $n$
increases with system size as $L^{1/2}$ whereas the optimal $\delta t$
behaves as $L^{-1/2}$. The obtained optimal values of $n$ were such that
$\bar C_M(1)$ still has the local minimum as a function of $\delta t$
mentioned above and which is near the absolute minimum of the energy.
For large $L$, the corresponding correlation times $\tau_M$ and $\tau_{\ch}$
turn out to be $L$--independent and are given by $\tau_M=2.5$,
$\tau_{\ch}=1.5$ approximately. The optimal acceptance probability
$p \approx 0.67$ and the product $n \delta t \approx 1$ are also
independent of the system size.

In summary, the computational effort both for the magnetization and the
energy behaves as $t_M, ~t_{\ch} \sim L^{1/2}$, corresponding to an
exponent $c=1/2$. This can be understood in the following way: In order
to keep the acceptance probability constant as we increase the system
size, $\delta t$ has to be varied as $L^{-d/4}$. Thus one needs to
increase $n$ as $L^{d/4}$ in order that the product $n \delta t$
appearing on the correlation function remains also unchanged as the
system size grows. The same picture was also seen to apply to a study
of the $\phi^4$ model performed with the standard HMC
method\cite{mehlig}. One should also note that an explicit
implementation of the method of Fourier acceleration needs the
calculation of Fourier transforms that involve an additional computer
effort of order $\log L$.

In conclusion, we have proposed a generalized version of HMC that is an
exact implementation of time-step matrix Langevin methods. We have
considered the Gaussian model as a test-case of the algorithm. The
optimal matrix $\ca$ is a diagonal matrix in Fourier space with diagonal
elements $\hat A_k=1/\omega_k$ (Fourier acceleration). This matrix
reduces completely CSD in the sense that the correlation times are mass
independent, $z=0$. We have proposed an approximation for the calculation
of the correlation functions of energy and magnetization that improves
as the system size gets larger. Using this approximation we have
discussed the optimization of the parameters of the algorithm. The
optimal computer effort grows as $L^{d/4}$ due to the decrease
of the acceptance probability with the system size and the need to keep
the trajectory length, $n \delta t$, constant at different system sizes.

Financial support from the Direcci\'on General de Investigaci\'on
Cient\'{\i}fica y T\'ecnica (Grant No. PB-92-0046, Spain) and JNICT
Portugal, are acknowledged. R. Toral thanks the warm hospitality at the
Departamento de F\'\i sica, Universidade de Aveiro, where this work was
carried out, and A. L. Ferreira thanks B. Mehlig for fruitful
discussions.

\vfill\eject

\vfill\eject
{\bf Figures Captions}
\vskip1truecm
{\bf Figure. 1.- } Comparison of $\bar C_M(1)$ as obtained from our analytical
approximation (continuous line) from  equation (\ref{CM}), for a system of size
$L=32$ and $n=4$ with simulation results (rhombi).
\vskip1truecm
{\bf Figure. 2.- } Comparison of $\bar C_{\ch}(1)$ (continuous line)
from equation (\ref{CH}),
with simulation results (rhombi) for the same system
size and number of leap-frog steps used in figure 1.
\vfill\eject

\clearpage
\begin{figure}[p]
\setlength{\unitlength}{0.240900pt}
\ifx\plotpoint\undefined\newsavebox{\plotpoint}\fi
\sbox{\plotpoint}{\rule[-0.175pt]{0.350pt}{0.350pt}}%

\end{figure}
\clearpage

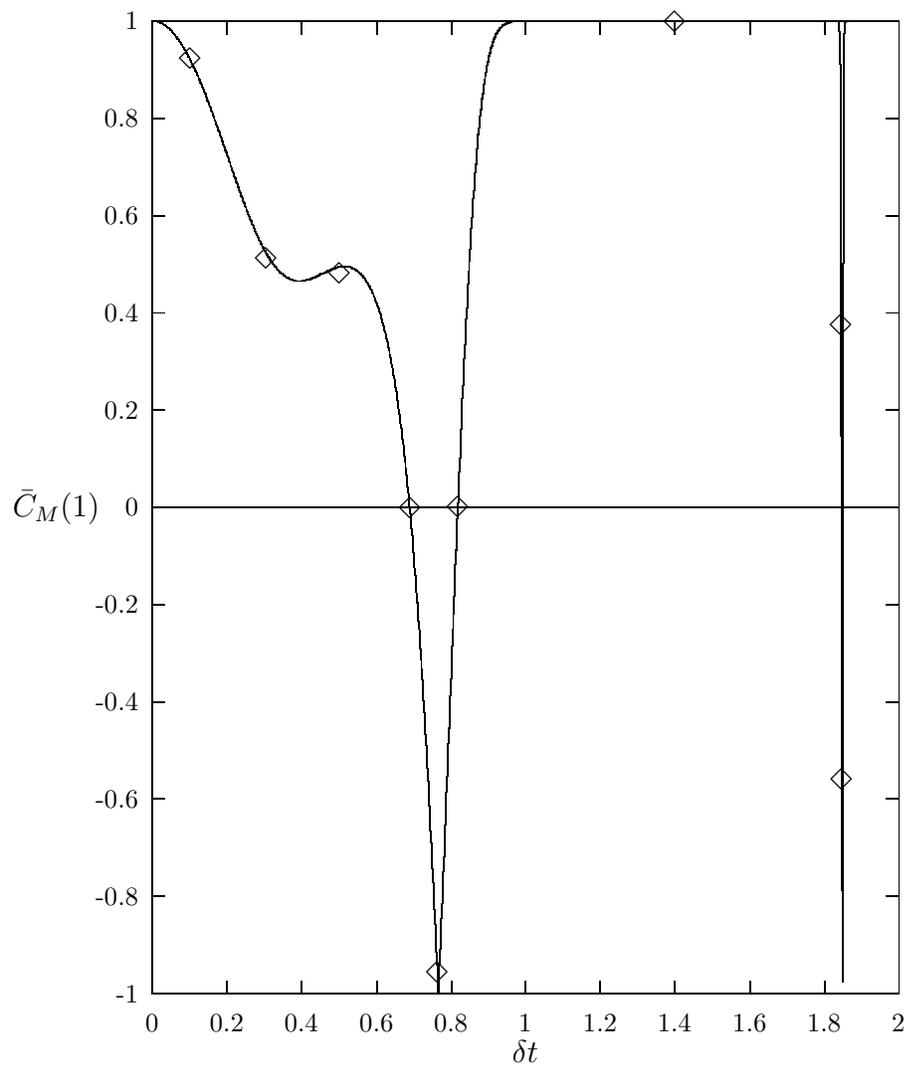
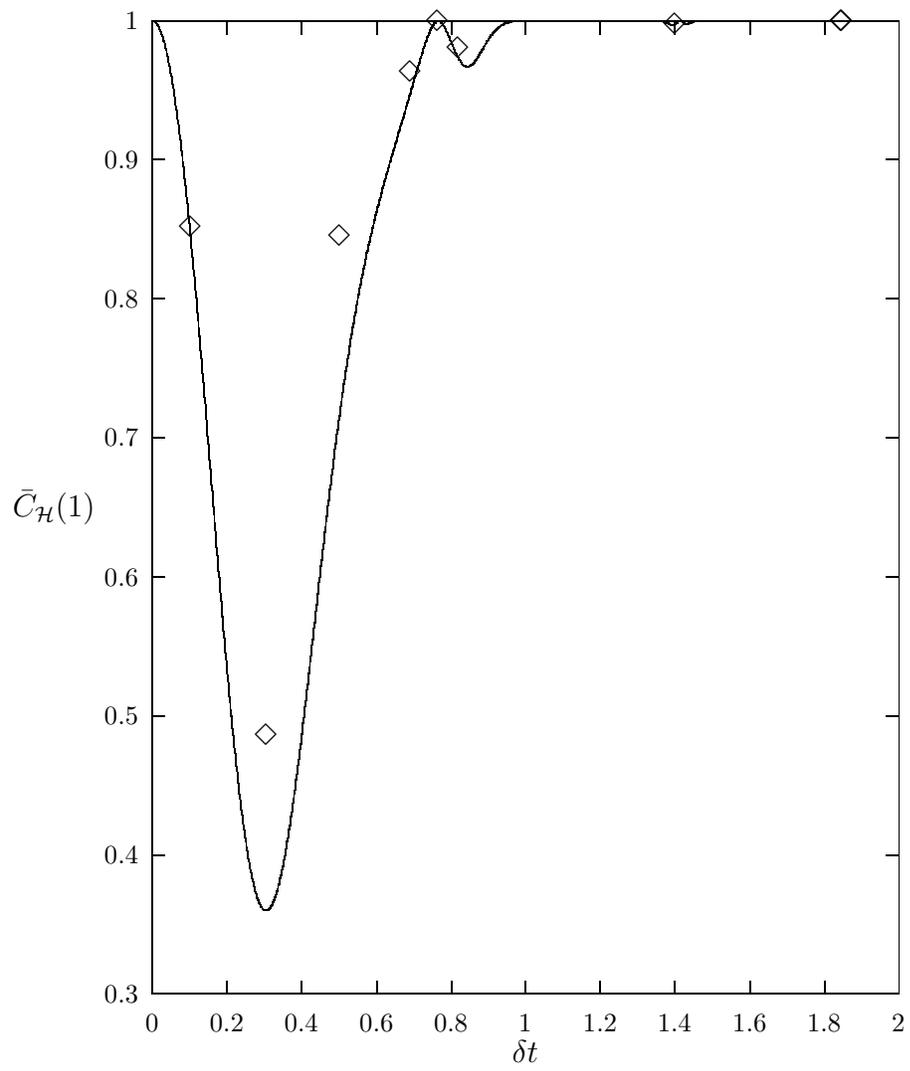
\end{document}